\documentclass[runningheads]{llncs}
\usepackage{amssymb}
\usepackage{amsmath}
\usepackage{psfrag}
\usepackage{epsfig}
\usepackage{cite}
\usepackage{graphics}
\usepackage{graphicx}
\usepackage{color}
\usepackage{subfigure}
\usepackage{multirow}
\usepackage{booktabs}
\usepackage{bbding}
\usepackage{pifont}
\begin{document}
\title{Unlocking the Potential of Early Epochs: Uncertainty-aware CT Metal Artifact Reduction}
\titlerunning{Uncertainty-aware CT MAR}

\author{
Xinquan Yang\inst{1} \and
Guanqun Zhou\inst{1,2} \and
Wei Sun\inst{3} \and
Youjian Zhang\inst{1} \and
Zhongya Wang\inst{1} \and
Jiahui He\inst{2} \and
Zhicheng Zhang\inst{1,2(}\Envelope\inst{)}
}

\institute{JancsiLab, JancsiTech, Hongkong, 999077, China
\email{\{yangxinquan,zhangyoujian,wangjie001,wangzhognya,zhouguanqun\}@jancsitech.net; zhangzhicheng13@mails.ucas.edu.cn}
\and
Shenzhen Institute of Advanced Technology, Chinese Academy of Science, 518055, Guangdong, China\\
\email{jh.he1@siat.ac.cn}
\and
University of Science and Technology of China, Anhui, 230026, China\\
\email{sunw@ustc.edu.cn}
}
%

%
\maketitle              
\begin{abstract}
In computed tomography (CT), the presence of metallic implants in patients often leads to disruptive artifacts in the reconstructed images, hindering accurate diagnosis. Recently, a large amount of supervised deep learning-based approaches have been proposed for metal artifact reduction (MAR). However, these methods neglect the influence of initial training weights. In this paper, we have discovered that the uncertainty image computed from the restoration result of initial training weights can effectively highlight high-frequency regions, including metal artifacts. This observation can be leveraged to assist the MAR network in removing metal artifacts. Therefore, we propose an uncertainty constraint (UC) loss that utilizes the uncertainty image as an adaptive weight to guide the MAR network to focus on the metal artifact region, leading to improved restoration. The proposed UC loss is designed to be a plug-and-play method, compatible with any MAR framework, and easily adoptable. To validate the effectiveness of the UC loss, we conduct extensive experiments on the public available Deeplesion and CLINIC-metal dataset. Experimental results demonstrate that the UC loss further optimizes the network training process and significantly improves the removal of metal artifacts.

\end{abstract}
\section{Introduction}

Computed tomography (CT) systems are indispensable tools for medical diagnosis and prognosis~\cite{jiang2023biology,jiang2022predicting}, and therapy planning~\cite{gardner2019modern,yang2024implantformer,yang2023tcslot,yang2023tceip,yang2024two,yang2024simplify} and guidance~\cite{simon2020role}. However, the presence of metallic implants in a patient’s body can compromise the reliability of X-ray projections, leading to pronounced star-shaped or streak artifacts in reconstructed CT images~\cite{katsura2018current}. Such artifacts significantly degrade image quality,
\begin{figure}
\centering
\includegraphics[width=1.0\linewidth]{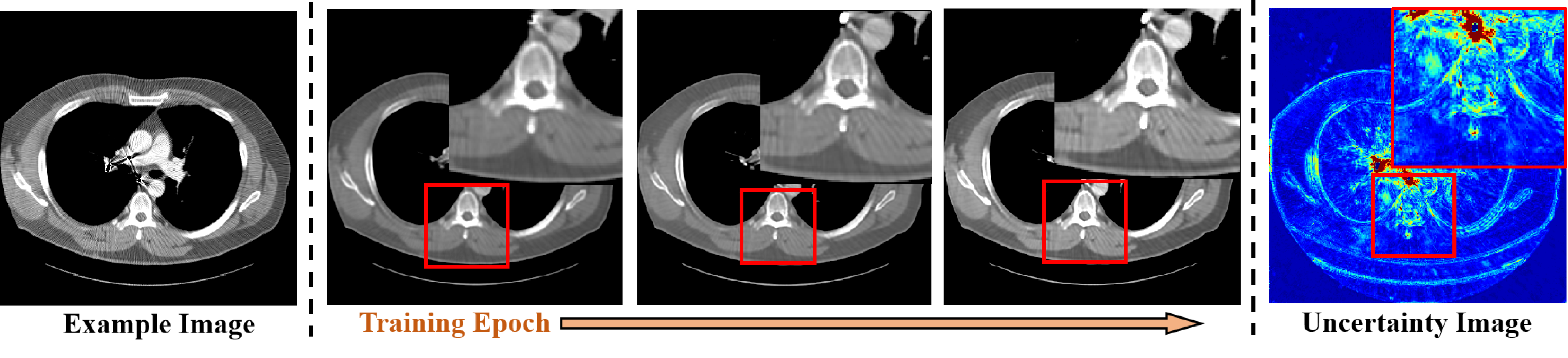}
\caption{Left: example image of CT with metal artifact. Middle: visualization of the restoration result of different training weight, in which the training epoch increases from left to right. Right: the uncertainty image calculated by the restoration results.} \label{fig1}
\end{figure}
undermining the diagnostic utility of CT scans~\cite{li2024quad}. Consequently, the development of effective metal artifact reduction (MAR) strategies has emerged as a critical issue, drawing heightened focus within the CT community.

With the advancement of deep learning, a considerable volume of methods has been designed to address the issue of MAR. These methods can be categorized into three types: (1) Sinogram domain-based methods: which restore metal-affected areas in the sinogram domain prior to the reconstruction process~\cite{liu2024unsupervised}. (2) Image domain-based methods: which directly learn the mapping from metal artifact images to clean images via deep neural networks~\cite{zhang2018convolutional,wang2023semimar,wang2021dicdnet,du2023deep}. (3) Dual domain-based methods: which use both the metal artifact image and the sinogram as inputs to achieve a better restoration result~\cite{choi2024dual,yu2020deep,yu2021metal,wang2021indudonet}.
Typically, the common practice for deep learning-based MAR methods is that we utilize a large amount of synthetic data to train the MAR network, and then apply the finely-tuned weights to test clinical data.
Although these methods have achieved some satisfactory results, their performance can be further improved, especially in the restoration of high-frequency texture.
According to previous research, neural networks tend to first restore low-frequency signals during the training process and then gradually reconstruct high-frequency signals~\cite{ulyanov2018deep}. This is consistent with our findings during the training of the MAR network, as illustrated in Fig.~\ref{fig1}, where we visualize the restoration results inferred from the initial training weights.
From these results, we can observe that as the training epochs increase, the texture information in the images becomes richer, but inevitably making metal artifacts more pronounced.
Based on this finding, we make most of the impact of initial training weights and use these restoration results to calculate an uncertainty image (see the right area of Fig.~\ref{fig1}), which highlights the high-frequency regions including metal artifacts and can effectively help the MAR network remove metal artifacts. This phenomenon inspires us to further exploit these restoration results.

\begin{figure}[htbp]
\centering
\includegraphics[width=1.0\linewidth]{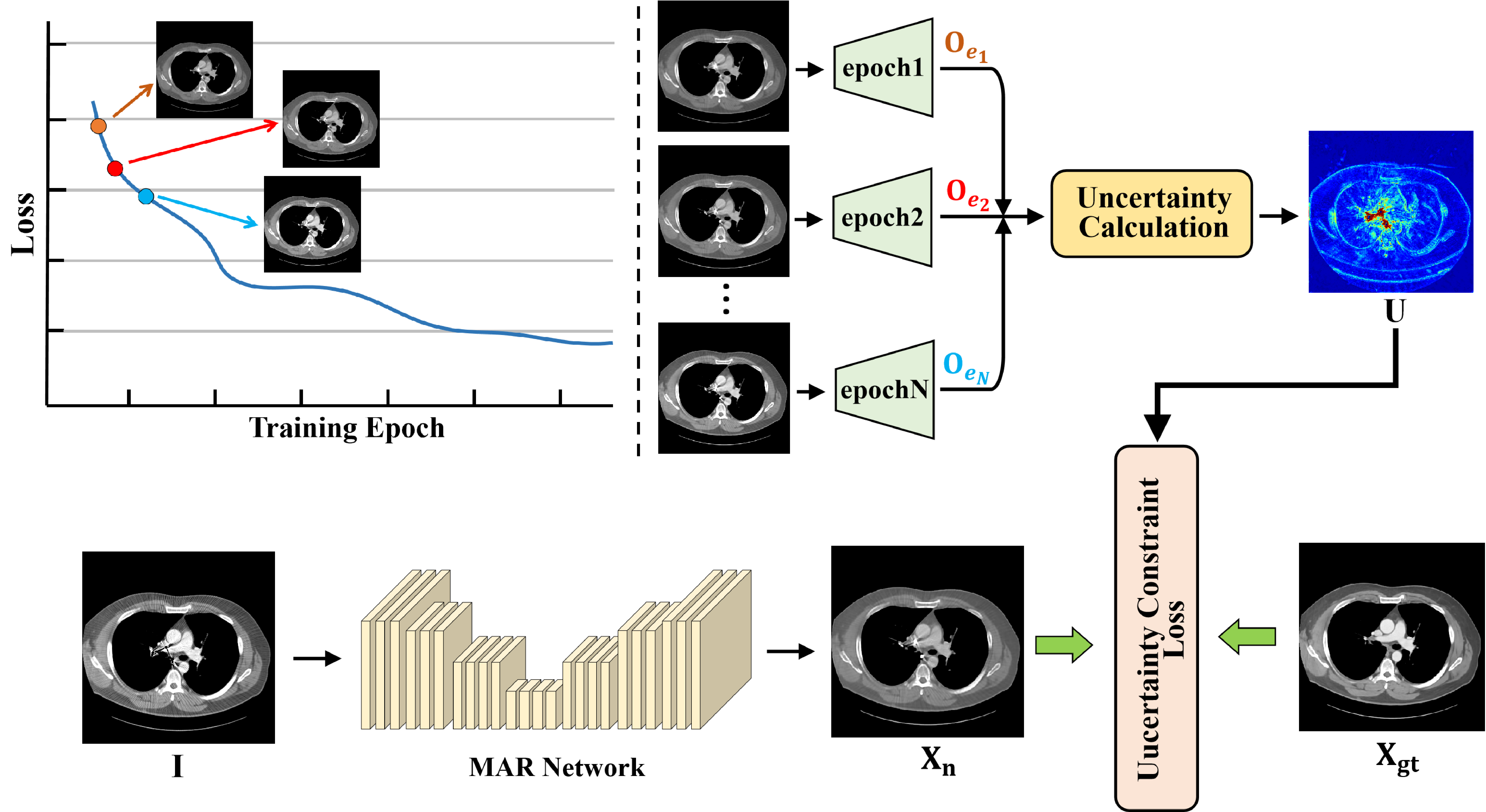}
\caption{The overview of the proposed uncertainty constraint loss.}
\label{fig_network}
\end{figure}

Motivated by the findings from Fig.~\ref{fig1}, in this study, we introduce an uncertainty constraint (UC) loss to guide the MAR network's attention toward regions with metal artifacts. As depicted in Fig.~\ref{fig_network}, our approach begins by training the MAR network for several epochs. We then apply the initial training weights to infer the input image and produce restoration results. Subsequently, these results are used to compute an uncertainty image, which, in turn, serves as an adaptive weight for the UC loss. This strategy enables the UC loss to target artifact removal. Notably, the UC loss is an elegant, plug-and-play tool that can be seamlessly integrated into any existing MAR network.
The key contributions of this paper are fourfold: (1) It pioneers the use of initial training weights to enhance the performance of MAR networks. (2) It introduces a novel UC loss that encourages the network to focus on metal artifact regions, leading to superior restoration performance. (3) The UC loss is plug-and-play and designed for ease of adoption, compatible with any MAR networks. (4) Comprehensive testings on both the Deeplesion and clinical datasets validate the the effectiveness of the proposed UC loss.

\section{Method}
\subsection{Metal Artifact in Initial Training}

Generally, a large amount of synthetic data is utilized to train the MAR network, which then be applied to test clinical data. However, under this paradigm, there is still room for further improvement in the metal artifact removal. From the visualization in Fig.~\ref{fig1}, we can see that as the training epochs increase, the texture information appears gradually, inevitably including metal artifacts. To this end, we use these restoration results derived from initial training weights to calculate an uncertainty image (see the right area of Fig.~\ref{fig1}) and then design an uncertainty constraint loss which can guide the MAR network to focus the high-frequency regions including metal artifacts. We introduce the UC loss in the next section.

\subsection{Uncertainty Constraint Loss}
As previously delineated, the restoration outcomes deduced from the initial training weights are employed to compute an uncertainty image. This image serves as an adaptive weight, directing the MAR network towards enhanced restoration performance. Consequently,
a plug-and-play uncertainty constraint loss is designed to refine the restoration process by quantitatively incorporating the uncertainty inherent in the initial predictions, thereby informing the subsequent optimization steps in a targeted manner.
As shown in Fig.~\ref{fig_network}, we first train the MAR network for a few epochs and the RMSE loss is used for supervision:

\begin{equation}
    \mathcal{L}_{RMSE} = \sqrt{\frac{1}{M} \sum_{m=1}^{M} (X_p - X_{gt})^2}
\end{equation}
where $X_{p}$ and $X_{gt}$ are the predicted result of MAR network and ground truth image, respectively.
Then, the initial training weights are employed to inference the training image, and the output images are used to calculate the uncertainty image:

\begin{equation}
\begin{array}{cccc}
     O_{e_1} = \textbf{MAR}_{e_1}(I)  \\
     O_{e_2} = \textbf{MAR}_{e_2}(I)  \\
     \textbf{...} \\
     O_{e_N} = \textbf{MAR}_{e_N}(I)
\end{array}
\end{equation}
where $I$ is the metal artifact-corrupted CT image. $O$ is the output of MAR network. $e_N$ denotes the training weight of the $N$-th epoch. After obtaining the restoration results, the uncertainty image $U$ can be calculated as follow:
\begin{equation}
    U = \textbf{Norm}(\textbf{std}(O_{e_1},...,O_{e_N}))
\end{equation}

Here, $\textbf{std}$ is the stands for the operation of standard deviation. $\textbf{Norm}$ is the operation of normalization, which scales the uncertainty image $U$ to a range where the minimum value is 0 and the maximum value is 1.
Then, we retrain the MAR network and use the proposed UC loss for supervision. The definition of UC loss is as follow:
\begin{equation}
    \mathcal{L}_{UC} = \sqrt{\frac{1}{M} \sum_{m=1}^{M} \{(X_p - X_{gt})\odot(1+U)\}^2}
\end{equation}
\section{Experiments and Results}
\subsection{Dataset}
We evaluate the proposed method with both simulated and clinic metal artifact-corrupted CT image. For the simulation data, we choose 1000 CT images from the DeepLeison dataset~\cite{yan2018deep} and follow~\cite{lin2019dudonet} to synthesize metal artifacts, in which 90 metal implants with different shapes and sizes are selected from the metal mask collection in~\cite{zhang2018convolutional}. Specifically, we randomly choose 800 CT images and 80 metal implants as the training set, and the remaining 200 CT images paired with 10 metal implants are set as test set. Both have no overlap. We use the public clinical dataset, \textit{i.e.}, CLINIC-metal~\cite{liu2021deep} to further evaluate the feasibility of the proposed method. The dataset includes 75 patients, and each patient has at least one artificial implant. We resize and process the clinical images using the same protocol to the synthesized dataset.

\subsection{Implementation Details}
The input image size of the MAR network is set to $416\times 416$ for network training and inference. We use a batch size of 4, Adam optimizer and a learning rate of $10^{-4}$ for the network training. To obtain the uncertainty image, we first train the MAR network for a few epochs and then stop the network training. Subsequently, we compute the uncertainty image to construct the UC loss for supervision. Total training epochs is 100 and the learning rate adjustment strategy uses cosine annealing, with a minimum learning rate of $10^{-6}$ and a period of 200. All the models are trained and tested on the platform of NVIDIA A40 GPU.

\subsection{Evaluation Metrics}
Peak signal-to-noise ratio (PSNR) and structured similarity index (SSIM) are employed to evaluate the performance on the simulated data. For the clinical data, due to the metal-free CT image is not available, we visualize the restoration result to demonstrate the effectiveness of the proposed method.


\subsection{Experiment Results on Synthetic Dataset}

\begin{table}[htbp]
\caption{Quantitative comparison of different methods on the synthetic dataset.}
\label{table_psnr_ssim}
\centering
\begin{tabular}{c|c|c|c}
\toprule
Method                     & UC loss & PSNR  & SSIM  \\ \hline
\multirow{2}{*}{UNet}      & \ding{55}       & 38.21 & 0.874 \\
                           & \Checkmark       & 41.33(8.16$\%\uparrow$) & 0.891(1.94$\%\uparrow$) \\ \hline
\multirow{2}{*}{IndudoNet} & \ding{55}       & 46.88 & 0.978 \\
                           & \Checkmark       & 48.67(3.82$\%\uparrow$) & 0.986(0.81$\%\uparrow$) \\ \hline
\multirow{2}{*}{MEPNet}    & \ding{55}       & 53.08 & 0.989 \\
                           & \Checkmark       & 54.71(3.07$\%\uparrow$) & 0.993(0.40$\%\uparrow$) \\ \hline
\multirow{2}{*}{DICDNet}   & \ding{55}       & 47.55 & 0.984 \\
                           & \Checkmark       & 50.05(5.25$\%\uparrow$) & 0.991(0.71$\%\uparrow$) \\
\bottomrule
\end{tabular}
\end{table}
\subsubsection{Quantitative Comparison on Different MAR Network.}
To further validate the performance of the UC loss, we compare two metrics, \textit{i.e.}, PSNR and SSIM, on the Deeplesion dataset. The experimental results are given in Table~\ref{table_psnr_ssim}. From the table, we can observe that the UC loss brings large improvement for the image domain-based methods than the dual domain-based ones. To be specific, Unet and DICDNet achieves 8.16\% and 5.25\% improvement, respectively, while IndudoNet and MEPNet only 3.82\% and 3.07\%.
This phenomenon is consistent with the design of UC loss, which computes the uncertainty image according to the restoration image, and the uncertainty image is used to supervise the restoration in the image domain. In terms of SSIM, Unet achieves the largest improvement (1.94\%) among all MAR networks, and the MEPnet achieves the best performance (0.993).

\begin{figure}
\centering
\includegraphics[width=0.85\linewidth]{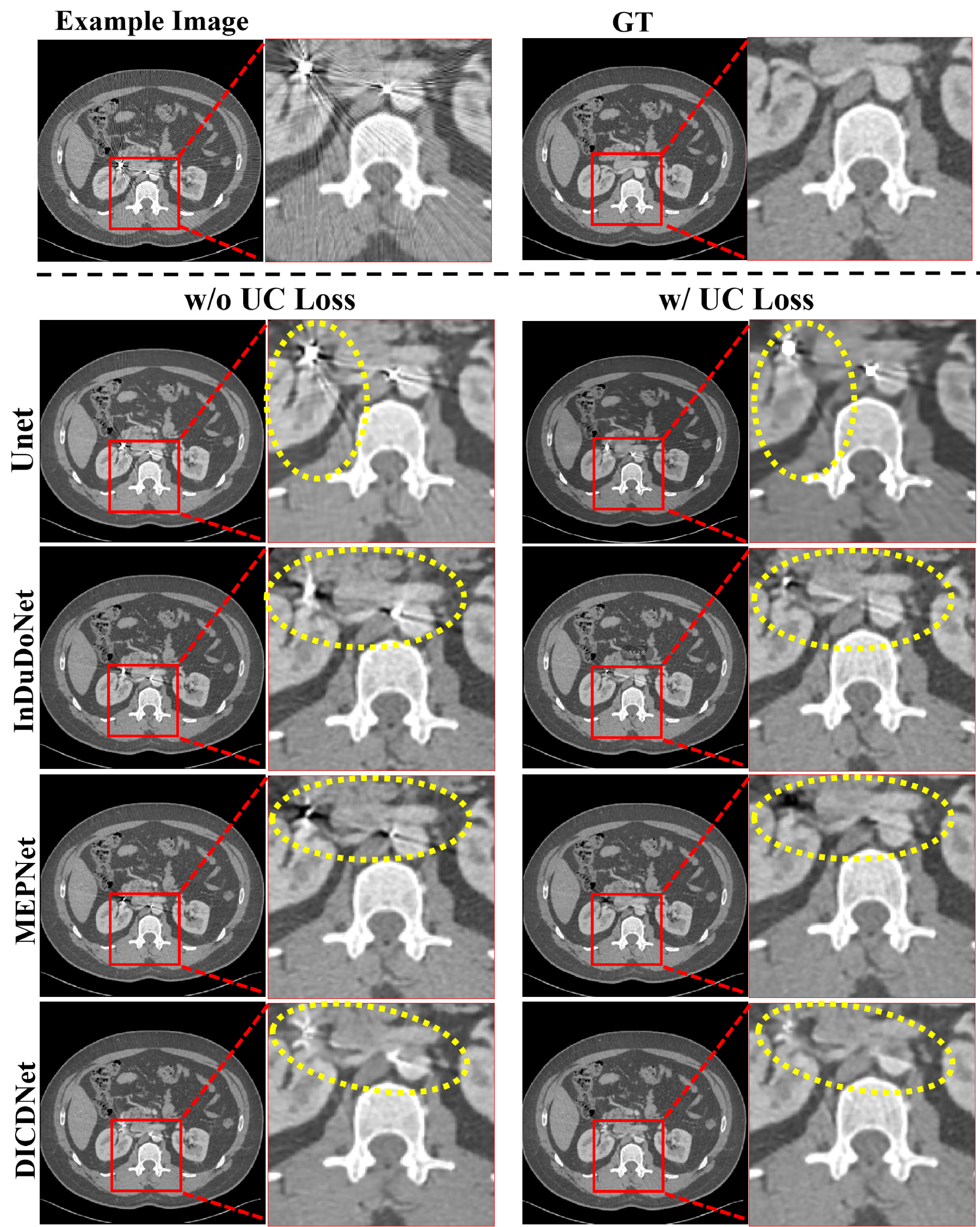}
\caption{Visual comparison of different MAR network with or without the UC loss for supervision on synthetic dataset. The red rectangle denotes the enlarged region.} \label{fig_syn}
\end{figure}

\subsubsection{Visual Comparison of Different MAR Network.}

In Fig.~\ref{fig_syn}, we present the restoration results of various MAR networks, both with and without the guidance of UC loss. As indicated in Fig.~\ref{fig_syn}, all methods were capable of diminishing the metal artifacts present in metal-corrupted CT images to varying extents, even in the absence of UC loss. Among these methods, DICDNet showed a relatively superior performance.
With the incorporation of UC loss, a marked improvement in the performance of all MAR approaches was observed. Notably, MEPNet's restoration results were the most closely aligned with the actual ground truth, attributed to MEPNet's inherent ability to effectively remove metal artifacts. On the other hand, the enhancement in visual quality for Unet, both before and after the introduction of UC loss, was particularly significant, with a visibly substantial reduction in metal artifacts. This evidence suggests that our approach serves as a plug-and-play strategy, which, when applied, can lead to considerable performance gains in methods like Unet that may initially have less satisfactory metal artifact removal capabilities.


\begin{figure}
\centering
\includegraphics[width=0.9\linewidth]{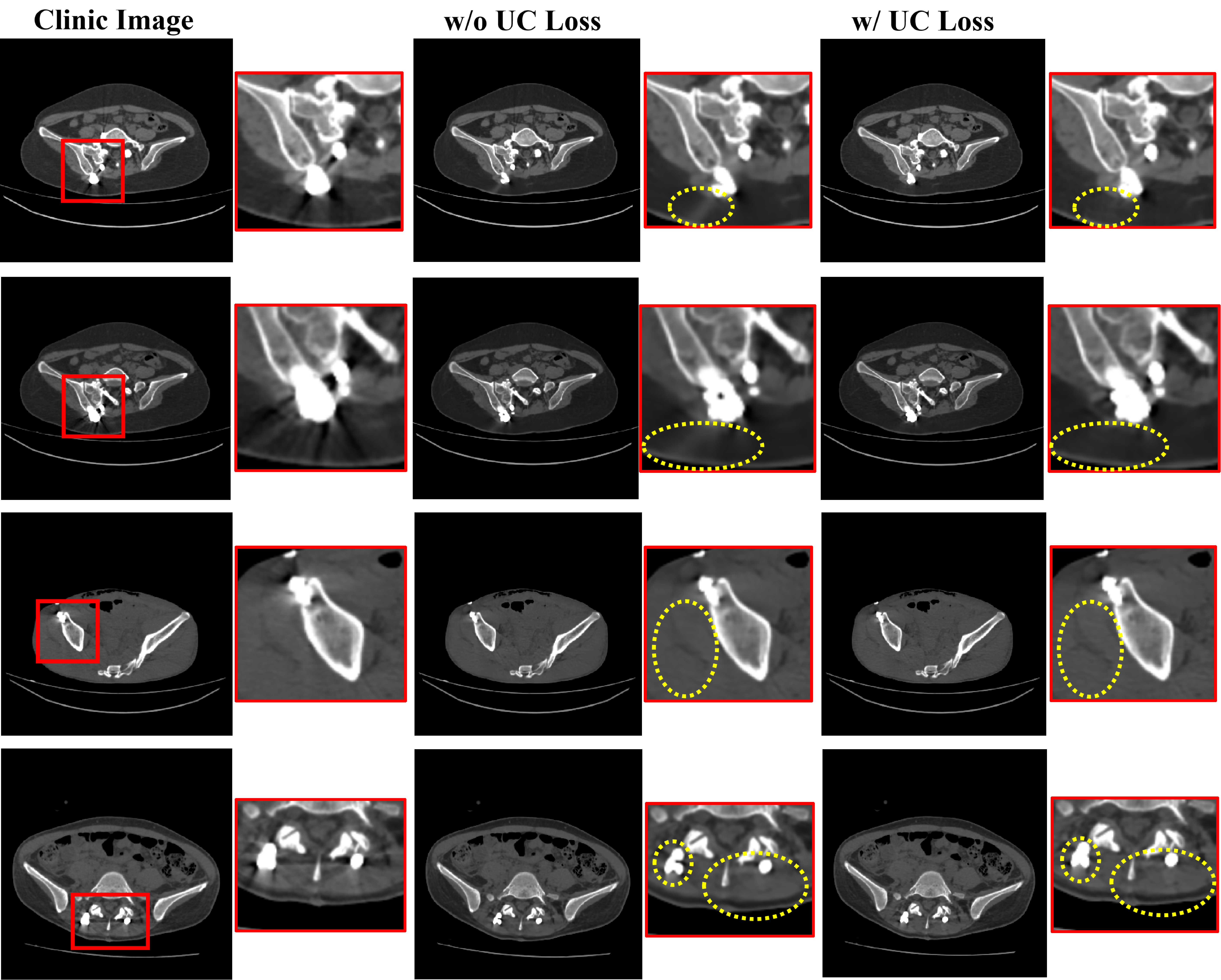}
\caption{Visual comparison of MAR network with or without the UC loss for supervision on clinic dataset. The red rectangle denotes the enlarged region.} \label{fig_clinic}
\end{figure}

\subsection{Experiment on Clinical Dataset}
To demonstrate the clinical effectiveness of the proposed UC loss, we visualize the restoration results on clinical dataset (CLINIC-metal) using MAR networks trained with or without the UC loss.
In real clinical scenario, the original sinogram data with metal artifacts are not available, and manually setting HU values to segment metals is not accurate. Therefore, we use the simplest image domain-based MAR network - Unet for visualization, which is a encoder-decoder architecture and only takes a single metal-corrupted CT image as input. The restoration results are shown in Fig.~\ref{fig_clinic}.

Examination of the first two rows in Figure~\ref{fig_clinic} reveals that the restoration effect of U-Net, in the absence of UC loss, is moderate, leaving a small quantity of residual metal artifacts apparent in the CT images. The incorporation of UC loss, as evidenced in these images, significantly diminishes the residual metal artifacts, rendering them virtually undetectable to the naked eye. Further analysis of the third row in Figure~\ref{fig_clinic} indicates that the application of UC loss enhances the definition of the CT images, with a notable increase in the restoration of textural details.
Finally, the last row in Figure~\ref{fig_clinic} demonstrates that the metal artifacts reconstructed using the original U-Net are incomplete, potentially leading to clinical misdiagnosis. Conversely, U-Net, when trained with UC loss, achieves a more comprehensive restoration of metal artifacts and yields images with markedly improved texture clarity. This suggests that the UC loss-augmented U-Net model may offer significant advantages in clinical settings, enhancing diagnostic accuracy.


\section{Conclusions}
In this paper, we propose an uncertainty constraint loss to address the restoration of highlighted metal artifact regions in the MAR network. This loss computes the uncertainty image using the restoration results obtained from the initial training weights. Our approach is designed as a plug-and-play method, seamlessly integrating into any existing MAR networks. Through extensive experiments on the Deeplesion and CLINIC-metal datasets, we demonstrate the effectiveness of the proposed UC loss in optimizing network training and significantly improving metal artifact removal.

\newpage
\bibliographystyle{splncs04}
\bibliography{ref}

\end{document}